# LARGE-APERTURE HIGH-FIELD NB$_3$SN MAGNETS FOR THE 2$^{ND}$ EIC INTERACTION REGION*

A.V. Zlobin#, I. Novitski, E. Barzi, Fermilab, Batavia, IL 60510, USA
B.R. Gamage, A. Seryi, Jefferson Lab, Newport News, VA, USA

*Abstract*

The design concept of the Electron Ion Collider (EIC), which is under construction at BNL, considers adding a 2$^{nd}$ Interaction Region (IR) and detector to the machine after completion of the present EIC project. Recent progress with development and fabrication of large-aperture high-field magnets based on the Nb$_3$Sn technology for the HL-LHC makes this technology interesting for the 2$^{nd}$ EIC IR. This paper summarizes the results of feasibility studies of large-aperture high-field Nb$_3$Sn dipoles and quadrupoles for the 2$^{nd}$ EIC IR.

## INTRODUCTION

The design concept of the Electron Ion Collider (EIC), which is under construction at BNL, considers adding a 2$^{nd}$ Interaction Region (IR) and detector to the machine after completion of the baseline EIC project [1]. They are to be installed in the IR8 region of the RHIC superconducting ring and will be used to cross-check the measurements and expand the EIC physics range [2]. To achieve these goals and provide enhanced performance and additional flexibility for the IR optics, the parameters of key IR magnets go beyond the capabilities of the traditional Nb-Ti magnet technology used in the 1$^{st}$ IR. Recent progress with the Nb$_3$Sn accelerator magnet technology [3], including the development of large-aperture high-field Nb$_3$Sn magnets for the HL-LHC [4], makes this technology interesting for use in the 2$^{nd}$ EIC IR. This paper summarizes the results of feasibility studies of the most challenging large-aperture high-field Nb$_3$Sn dipoles and quadrupoles for the 2$^{nd}$ IR. These studies included the aperture and field ranges, operation margins, and stress management in brittle Nb$_3$Sn coils.

## 2$^{ND}$ IR LAYOUT AND MAGNET PARAMETERS

The baseline design of the EIC IR makes use of the Nb-Ti magnet technology [1]. The 2$^{nd}$ IR is planned to provide complementary measurements to the primary IR by exploring additional parameter space of the scattered particles [1], [2]. The magnets in 2$^{nd}$ IR require larger apertures and higher field gradients to provide an optimal second focus configuration and wide acceptances for both charged and neutral scattered particles. Using Nb$_3$Sn magnets, two final focusing quadrupoles (FFQs) are sufficient to perform the functions of four Nb-Ti FFQs.

Two IR layout options are considered for the 2$^{nd}$ IR (see Figure 1). Option 1 includes two Nb$_3$Sn quadrupoles and two dipoles with correctors. Option 2 uses three FFQs and is better optimized for lower energies. In this layout, at higher energies the three magnets are operated as a doublet whereas at lower energies they work as a triplet to better accommodate the beam envelope and provide a better acceptance [5].

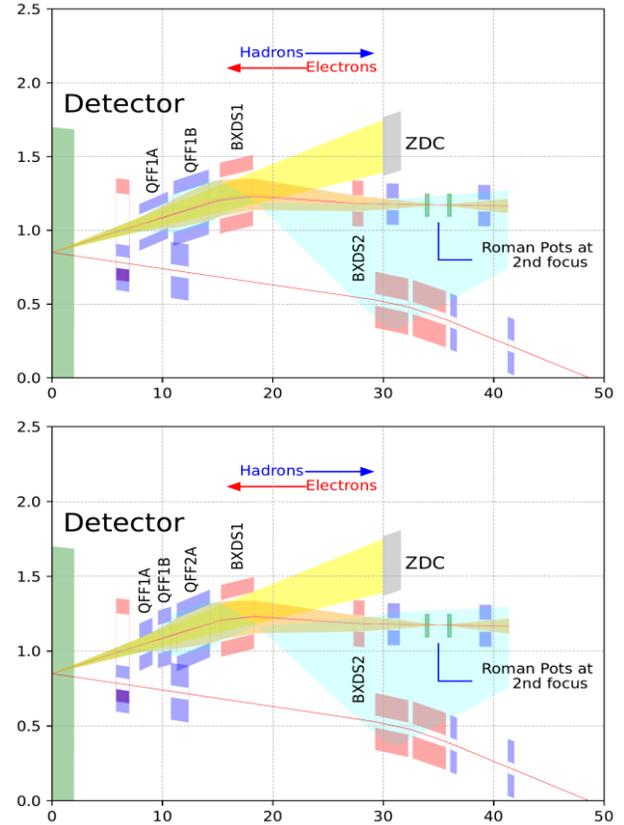

Figure 1: IR8 forward with Nb$_3$Sn magnets option 1 (top) and option 2 (bottom) [5].

Table 1. Nb$_3$Sn magnet parameters for IR options 1 and 2

| Magnet | Length [m] | Aperture [mm] | Field G [T/m] | B field [T] |
|---|---|---|---|---|
| QFF1A | 2.6*/1.2** | 173*/119** | -106.4*/155** | – |
| QFF2A | 3.2*/1.2** | 256*/190** | 71.9*/-65.6** | – |
| QFF2B | 3** | 258** | 71.6** | – |
| BXDS1 | 3 | 300*/280** | – | 8.6 |
| BXDS2 | 1 | 110 | – | -3.7 |

\* - option 1; \*\* - option 2.

The main parameters of the key magnets for the 2$^{nd}$ IR are summarized in Table 1. The magnetic field and the aperture of BXDS2 are close to the Nb-Ti D1 with 150 mm aperture and 6 T field, and those of QFF1A are close to the Nb$_3$Sn MQXF in HL-LHC [4]. This paper focuses on the Nb$_3$Sn IR quadrupole and dipole which could represent QFF2A/B and BXDS1.

___
* Work supported by Fermi Research Alliance, LLC, under contract No. DE-AC02-07CH11359 with the U.S. DOE and by Jefferson Science Associates, LLC under contract No. DE-AC05-06OR23177.
# zlobin@fnal.gov

# IR MAGNET DESIGNS AND PARAMETERS

Designs of the proposed IR quadrupole and dipole are based on $Nb_3Sn$ Rutherford cable developed and used in $Nb_3Sn$ dipole models fabricated and tested at Fermilab [6]. The cable is 15.1 mm wide, has a keystoned cross-section with the keystone angle of 0.805 degree, and mid-thickness of 1.319 mm. The cable has 40 $Nb_3Sn$ strands each 0.7 mm in diameter, with a Cu/non-Cu ratio of 1.13, and $J_c$ at 12 T and 4.2 K of 2500 A/mm$^2$. The strand and cable cross-sections are shown in Figure 2.

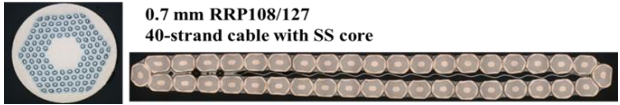

Figure 2: Cross-sections of $Nb_3Sn$ strands and cables used in this study.

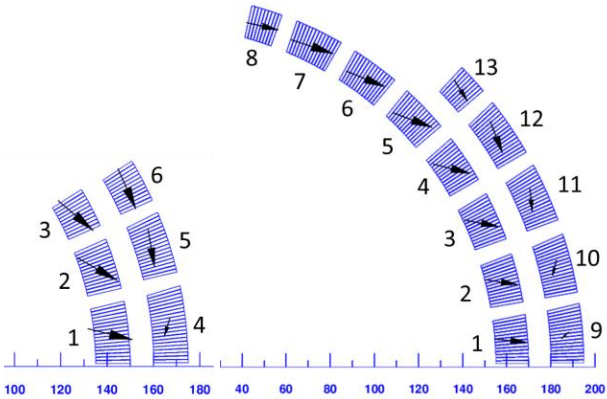

Figure 3: Cross-sections of one octant of IR quadrupole (left) and one quadrant of IR dipole (right) coil with block numbering and Lorentz force vectors in coil blocks.

Apertures of the IR quadrupole and dipole coils were increased to 270 mm and 310 mm respectively to take into account a 3-mm thick beam pipe and a 3-mm wide annular channel for the coil cooling by liquid Helium. The coil magnetic calculations were performed using *ROXIE* [7] for a cylindrical iron yoke with inner radius of 180 mm for IRQ and 200 mm for IRD, outer yoke radius of 500 mm for both magnets, and real iron $B(H)$ curve.

Preliminary optimized cross-sections of the IR quadrupole and dipole coils are shown in Figure 3. Coil turns in both coils are grouped into blocks separated by radial (interlayer) and azimuthal spacers. The role of these spacers will be discussed later. The coil blocks in each coil are numbered and the arrows in the coil blocks represent the relative value and direction of the Lorents forces in each block.

The optimized cross-sections of the quadrupole and dipole coils with the field uniformity diagram in the aperture and the field distributions in the coil blocks at a current of 12.5 kA, are shown in Figure 4. The good field quality areas, where relative field variations with respect to the main field component are smaller than 3·10$^{-4}$, are close to ~160 mm in diameter in both magnets (dark-blue areas in the aperture in Figure 4).

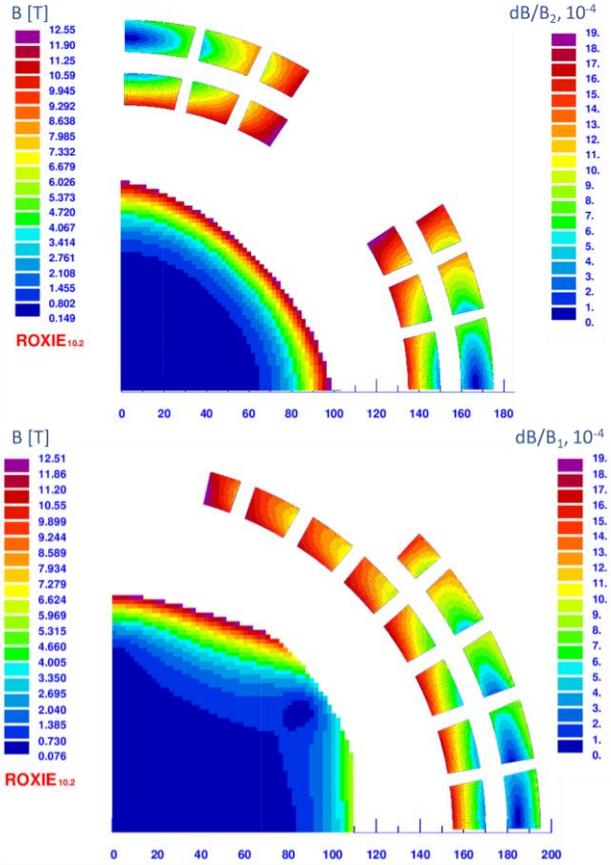

Figure 4: Field quality diagram in the aperture and field distribution in the quadrupole (top) and dipole (bottom) coil blocks at 12.5 kA current.

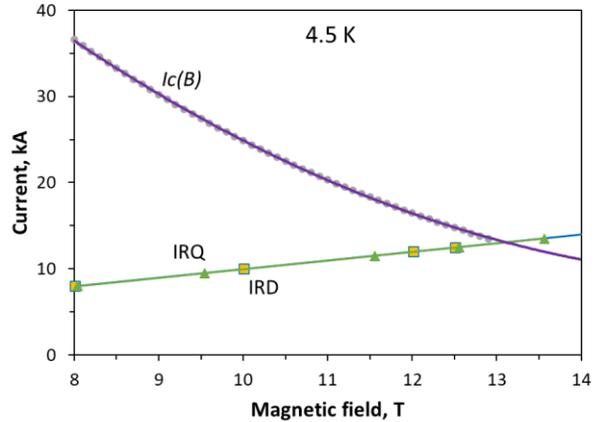

Figure 5: $I_c(B)$ curves for the 40-strand $Nb_3Sn$ cable measured at 4.5 K, and IR quadrupole QFFDS01B and dipole BXDS01A load lines.

Figure 5 shows the $I_c(B)$ dependence of the 40-strand $Nb_3Sn$ cable at 4.5 K and the load lines of IR quadrupole (IRQ) and dipole (IRD) coils. The main magnet parameters are summarized in Table 2. Since the magnet load lines are very close, the calculated field limit at the coil maximum current of ~12.5 kA is 12.5 T for both magnets. This corresponds to a field gradient in the IRQ of 78 T/m, and to a dipole field in the IRD of 10.5 T.

Table 2. IR magnet parameters

| Parameter | IRQ | IRD |
|---|---|---|
| Coil ID, mm | 270 | 310 |
| Iron yoke ID, mm | 360 | 400 |
| Coil current, kA | 12.5 | 12.5 |
| Coil field, T | 12.55 | 12.52 |
| Bore field/Field gradient, T/T/m | 77.96 | 10.50 |
| Margin wrt nominal value, % | 8.4 | 22.1 |

One can see that the $Nb_3Sn$ IRD provides significant margin of ~22% with respect to the nominal value shown in Table 1. If this large margin is not needed, the magnet nominal operation field could be increased by reducing simultaneously the dipole coil length, or the magnet coil width could be reduced by using a narrower $Nb_3Sn$ cable. The IRQ margin is ~8%. If necessary, it could be increased by adding turns to the coil or by using higher-$J_c$ or wider cable. Magnet operation at 1.9 K in superfluid Helium will increase the magnet operation margin by ~10%.

## COIL STRESS MANAGEMENT

The Lorentz forces in the blocks, shown in Figure 3, may cause high mechanical stresses in the coil and large coil block displacements with respect to the design positions. Since $Nb_3Sn$ superconductor is brittle and may degrade or even lose its superconducting properties at stresses above 150 MPa, the stresses and displacements need to be kept within acceptable limits. In large-aperture high-field magnets this is achieved by using a special stress management by winding the coil blocks into a special stress management structure [8].

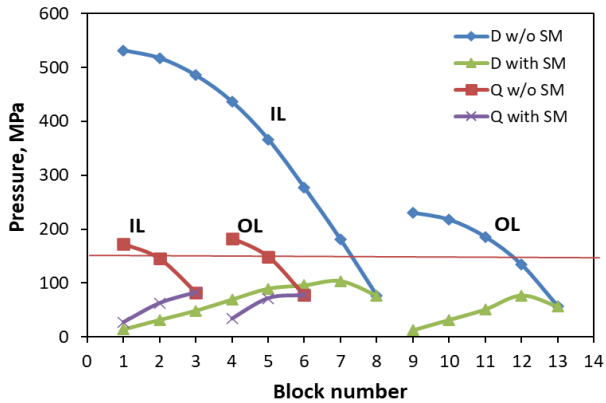

Figure 6: Variation of block average azimuthal stress in dipole and quadrupole coils without and with stress management. Block numbering is shown in Figure 3.

The calculated variation of block average azimuthal stress in the IR dipole and quadrupole coils at a current of 12.5 kA without and with stress management is plotted in Figure 6. The block numbering is shown in Figure 3. The horizontal line in the plot shows the stress limit of 150 MPa used for $Nb_3Sn$ superconductor. One can see that the stress level in the IR quarupole and dipole coils without stress management ecceeds the superconductor stress limit. Taking into acoount the large field variations in the coil blocks, the maximum stress in the blocks is even larger.

Thus, stress management is needed to keep the coil stress at the acceptable level for brittle $Nb_3Sn$ superconductor. This is obtained by filling the radial and azimuthal spaces between the blocks with a strong metallic structure [8].

## CONCLUSION

Conceptual designs of the large-aperture high-field quadrupole and dipole for the 2$^{nd}$ EIC IR based on $Nb_3Sn$ superconductor have been generated and analyzed. Preliminary analysis shows that the nominal values of field and field gradient, including margins, are achievable for the nominal apertures in 2$^{nd}$ IR using two-layer shell-type coil designs, state-of-the-art $Nb_3Sn$ technology and operation at 4.5 K. Further magnet design optimization, including field quality and operation margins, is possible and will be done during the engineering design phase.

Due to the large apertures and high fields in the coil, the stress level due to Lorentz forces in both magnets is substantial. It significantly exceeds the admissible stress for brittle $Nb_3Sn$ superconductor. The reduction of the coil stress to an acceptable level can be achieved using a stress management approach by splitting the magnet coils into blocks and winding them into special strong coil structures. The practical development and demonstration of this stress management technology for shell-type coils is in progress at Fermilab in the framework of the US Magnet Development Program [9]. The first experimental results are expected this year from prototypes. The design, materials and parameters of the stress management structure for the 2$^{nd}$ IR magnets will be also selected and optimized in the magnet engineering design phase.